\documentclass[prl,twocolumn,aps,amsmath,amssymb,showpacs]{revtex4}

\usepackage{t1enc}
\usepackage[final]{graphicx}
\usepackage{graphicx}
\usepackage{epsfig}

\begin{document}

\title{Topology, phase transitions and the spherical model}
\author{Sebastián Risau-Gusman\footnote{current address: Departamento de F\'{\i}sica,
            CNEA - Centro Atómico Bariloche
 (R8402AGP) San Carlos de Bariloche, Río Negro, Argentina}
}
\email{srisau@cab.cnea.gov.ar}
\affiliation{Departamento de F\'{\i}sica,
             Universidade Federal do Rio Grande do Sul,
             CP 15051, 91501-970, Porto Alegre, Brazil}
\author{Ana Carolina Ribeiro-Teixeira}
\email{anacarol@if.ufrgs.br}
\affiliation{Departamento de F\'{\i}sica,
             Universidade Federal do Rio Grande do Sul,
             CP 15051, 91501-970, Porto Alegre, Brazil}
\author{Daniel A. Stariolo}
\email{stariolo@if.ufrgs.br}
\altaffiliation{Research Associate of the Abdus Salam International 
Centre
for Theoretical Physics, Trieste, Italy}
\affiliation{Departamento de F\'{\i}sica,
             Universidade Federal do Rio Grande do Sul,
             CP 15051, 91501-970, Porto Alegre, Brazil}
\date{\today}

%%%%%%%%%%%%%%%%%%%%%%%%%%%%%%%%%%%%%%%%%%%%%%%%%%%%%%%%%%%%%%%%%%%%%

\begin{abstract}
The Topological Hypothesis states that phase transitions should be related to changes 
in the topology of configuration space. The {\it necessity} of such changes has 
already been demonstrated. We characterize exactly  
the topology of the configuration space of the short range Berlin-Kac spherical model, 
for spins lying in 
hypercubic lattices of dimension $d$. We find 
a continuum of changes in the topology and also a finite number of discontinuities in 
some topological functions. 
We show however that these discontinuities {\it do not} coincide with the 
phase transitions which happen for $d \geq 3$, and conversely, 
that no topological discontinuity can be associated to them.
This is the first short range, confining potential for which 
the existence of special topological changes are
shown {\em not to be sufficient} to infer the occurrence of a phase 
transition. 
\end{abstract}


\maketitle

Phase transitions (PTs) remain one of the most intriguing and 
interesting phenomena in physics. Mathematically, a PT is 
signaled by the loss of analyticity of some thermodynamic 
function~\cite{YL} in the thermodynamic limit.

Recently, a new characterization of PTs has been proposed, that conjectures that 
"at their deepest level PTs of a system are due to a change of the topology of
suitable submanifolds in its configuration space"~\cite{CPC2}.
%PTs "would at a deeper level be related to a particular change in the topology of
% the configuration space of the system"~\cite{Review}. 
This is known as the Topological Hypothesis (TH)~\cite{Review}. 
In this new method one studies 
the topology of the configuration space $\Gamma$ of the potential energy 
$V({\bf x})$ of a system with $N$ degrees of freedom, determining the changes 
that take place in the manifolds $M_v=\{ {\bf x} \in \Gamma : V({\bf 
x})/N<v \}$ as the parameter $v$ is increased.
A topological transition (TT) is 
said to take place at $c$ if $M_{c-\epsilon}$ and 
$M_{c+\epsilon}$ are not homeomorphic. The idea is that somehow TTs 
may be related to PTs. 

The {\it necessity} of TTs at a phase transition point has been demonstrated for short 
ranged, confining models~\cite{FP}. In the XY model~\cite{CPC2} TTs are found 
both in the mean field (MF) and unidimensional short range (SR) versions, 
whereas a PT is present only in the
MF case. This led to a refinement of the TH: only sufficiently 
``strong'' TTs would be able to induce a PT. It was found that in the MF 
version a {\it macroscopic} change of the Euler characteristic happens at 
exactly the same point $v_c$ where a PT appears. Several other models seem to be 
in agreement with this behavior~\cite{G1}.
But recently it was proved for a nonconfining potential that no topological criterion 
seems to be {\it sufficient} to induce a PT~\cite{G2}. We show below that the same happens for the spherical model, which is a confining, short ranged potential.

In the spherical model it has been found that there is a 
direct correlation between the TT and the PT, in its mean field 
version~\cite{RS}. 
Interestingly, in the case of nonvanishing external field there is no 
PT, but the configuration space displays a TT at energies that cannot 
be thermodynamically reached.

In this letter we study the original Berlin-Kac spherical model~\cite{BK} for 
spins placed on a 
$d$-dimensional lattice, interacting with their first neighbors. Using 
tools from topology theory 
we were able, for the case of vanishing field, to determine its 
topology 
{\it exactly} (up to homology). We show that the PT 
occurring for $d \geq 3$ cannot be related to {\it any} discontinuity in the 
homology of the manifolds at $v_c$. 
For nonvanishing field we cannot characterize the topology completely 
for all $v$, but show that a very abrupt change 
in the 
topology happens that does not have a corresponding PT. At variance 
with the MF version, the value of $v$ at which this topological change occurs
{\it is thermodynamically accessible}.

The spherical model is defined by a set of $N$ spins {$\epsilon_i$} 
lying on a $d$ dimensional hypercubic lattice and interacting 
through the potential 
$V=-\frac{1}{2} \sum_{<ij>} J_{ij}  \, \epsilon_i \epsilon_j - H \sum_i 
\epsilon_i$
where $J_{ij}=J$ gives the strength of the interaction between 
nearest-neighbor spins $i$ and $j$, and $H$ is an external field. The spin 
variables
are real and constrained to lie on the sphere $\mathbb{S}^{N-1}$ (i. e. 
$\sum_i \epsilon_i^2=N$). 
Periodic boundary conditions are imposed on the lattice.

In~\cite{BK} it is shown that, at zero 
field, a continuous PT appears at a critical temperature 
$T_c(d)$ for $d\geq3$, which is a strictly increasing function of $d$ 
(see Table 1). On the other hand, no PT is possible in an 
external field. 

As in previous works, the thermodynamic function we use to 
relate the statistical mechanical and topological approaches is the 
{\em average potential energy per particle} $\langle v \rangle $. 
%Straightforwardly generalizing the results of~\cite{BK} we obtain $T_c$ and  $\langle v_c \rangle $ for all dimensions (see Table 1). 
Although the specific details 
of $\langle v \rangle$ depend on $d$, some 
features are common to all hypercubic lattices: $\langle v \rangle  \rightarrow 0$ for
$T\to \infty$ and $\langle v \rangle  \rightarrow -d$ (its lower bound) when 
$T\to 0$.

\begin{table}[ht]
\begin{center}
\begin{tabular}{|c|c|c|}        \hline
{\, $d$ \, }     &{\,$kT_c/J$\,}       &{\,$\langle v_c \rangle /J$\,}
\\ \hline
3       &3.9573 &-1.0216 
\\ \hline
4       &6.4537 &-0.7728 
\\ \hline
5 &8.6468 &-0.6759
\\ \hline
6 &10.7411  &-0.6283
\\ \hline
\end{tabular}
\end{center}
\caption{Critical temperatures $T_c$ and mean potential 
energies per particle $\langle v_c \rangle$ for hypercubic lattices in $d$ dimensions. 
Values obtained from analytical expressions in~\cite{BK}.}
\label{tab:table1}
\end{table}

In the topological approach one looks for changes in the topology of $M_v$ as $v$ is 
increased. A topological change happens at a certain value $v_{T}$ if 
the manifolds $M_{v_{T}-\epsilon}$ and $M_{v_{T}+\epsilon}$ are not 
homeomorphic~\cite{Hatcher} for arbitrarily small $\epsilon$.
To make a connection with statistical mechanics Casetti et al.~\cite{Review} 
proposed the nontrivial ansatz that, at the phase transition, $v_T$ can 
be identified 
with $\langle v_c \rangle $, the thermodynamical average critical potential 
energy per particle. 
To study the topology of the configuration space of the spherical model 
it is 
most convenient to write the potential using the coordinates $x_i$ that 
diagonalize the 
interaction matrix through an orthogonal transformation:
\begin{equation}
V=-\frac{1}{2}\sum_{i=1}^N \lambda_i x_i^2-\sqrt{N}x_1 H
\label{potential}
\end{equation}
\noindent where we set $J=1$, and $\lambda_i$ ($i=1,\cdots,N$) are the 
eigenvalues of the 
interaction matrix, ordered from largest to smallest.  We 
define the {\it sets} $C_j$, $j=0, \cdots ,{\hat 
N}$, where {$\hat N+1$} is the number of {\it distinct} eigenvalues. 
$C_j$ is the set containing the indices of the eigenvalues that have 
the 
$(j+1)$-th largest value. Therefore, $|C_j|$ gives the degeneracy 
associated 
to the $(j+1)$-th largest eigenvalue. The 
Frobenius-Perron theorem ensures that the largest eigenvalue is not 
degenerated, i. e. $C_0=\{1\}$.

The critical points of this potential on the sphere 
$\Gamma= \mathbb{S}^{N-1}=\{ {\bf x} \in \mathbb{R}^N : \sum_{j=1}^N x_j^2=N \}$ are found using 
Lagrange multipliers. Along with the spherical constraint, the critical 
point equations are:
\begin{eqnarray}
x_1(2\mu + \lambda_1)+\sqrt{N}H &=& 0 \label{eq.critp} \\
x_i(2\mu + \lambda_i )&=& 0 \,\,\,\,\,i=2, \cdots ,N \nonumber
\end{eqnarray}
\noindent where $\mu$ is the Lagrangian multiplier that results from 
enforcing the spherical constraint. From these equations and 
Eq.~(\ref{potential}) ${\hat N}+1$ critical 
values of $v$ are obtained, denoted $v_k=-\lambda_l/2$, with $l\in C_k$, and
$k=0, \cdots ,{\hat N}$ (ordered from smallest to largest). 
Notice that the degeneracy of the eigenvalues causes that the 
corresponding critical points be in fact critical 
{\it submanifolds}. This implies 
that in the directions tangent to the critical submanifolds
the Hessian vanishes, which in turn implies that the potential is not a 
proper 
Morse function. Nevertheless, using Bott's extension of Morse theory the Euler 
characteristic can be found exactly.
%More precisely, the Hessian has 
%$\sum_{i=0}^{j-1} |C_i|$ negative 
%eigenvalues when restricted to the submanifold 
%normal to the $j$th critical submanifold.
However, profiting from the symmetries of the spherical model 
we took a more direct route to study its topology. As we show below, for vanishing 
external field it is possible to 
characterize {\it completely} the topology of the $M_v$, by explicitly giving 
the values of {\it all} the Betti numbers of the manifolds.
%Within Morse theory one can only obtain the alternate sum of the Betti 
%numbers (i.e. the Euler characteristic), or bounds for them~\cite{Milnor}.

For $H=0$ the critical manifolds 
$\Sigma_{v_j}$, $j=0, \cdots ,{\hat N}$, are given by 
$\Sigma_{v_j}=\{ {\bf x} \in \Gamma: \sum_{i \in 
C_j} x_i^2=N\}$ (see eq. (\ref{eq.critp})). These are 
(hyper)spheres whose dimension is given by the 
degeneracy of the corresponding eigenvalues.
To understand the nature of the topological change that 
happens at the 
critical values of $v$ it is necessary to know the topology of the 
$M_v$ for $v$ between two critical values. 
We show below that in the interval $(v_{j},v_{j+1})$ all the manifolds 
$M_v$ are 
{\it homotopy equivalent} to 
$\mathbb{S}^{D-1}$, where $D$ is the number of
eigenvalues larger than $-2v$. In fact we 
prove that $\mathbb{S}^{D-1}$ 
is a deformation retract of $M_v$, which in turn implies their 
homotopy equivalence~\cite{Hatcher}.

A submanifold $S \subset M$ is a deformation retract of a manifold $M$ if there 
exists a series of maps $f^{\nu}: M \rightarrow M$ with $\nu \in 
[0,1]$, 
such that $f^{0}=I$, $f^{1}(M)=S$ and $f^{\nu}|_S=I$ for all $\nu$. The 
map when considered as $f:M \times [0,1] \rightarrow M$ must be 
continuous.
Let us take $v \in (v_{j},v_{j+1})$. The deformation 
retract that takes the manifold $M_v$ onto its submanifold 
$\mathbb{S}^{D-1}=\{ {\bf x} \in M_v: \sum_{i=1}^D x_i^2=N \}$ is given by 
${\bf x}(\nu)=(f^{\nu}_1 
({\bf x}), \cdots, f^{\nu}_N (\bf x))$ with

\begin{equation}
f^{\nu}_i({\bf x}) = \left\{ \begin{array}{ll} 
 x_i \sqrt{1+\nu \sum_{k=D+1}^N x_k^2 / \sum_{k=1}^D x_k^2} & \mbox{for   
}i \leq D \\
 & \\
x_i \sqrt{1-\nu} & \mbox{for   }i > D
\end{array}
            \right.
\label{eq.retract}
\end{equation}

This map can easily be shown to be continuous at all points $\mathbf{x} \in M_v$.
The properties for $\nu=0$ and $\nu=1$ are evidently fulfilled. 
It is also easy to see that the retraction $f^{\nu}$ does not map any points outside $M_v$, since the image points always lie on the sphere 
$\mathbb{S}^{N-1}$, and their potential energy does not exceed $v$.
%It can 
%also be seen that no points are mapped outside $M_v$, as follows. It is 
%easy to see 
%that the image points always lie on the sphere $\mathbb{S}^{N-1}$, 
%but it must also be checked that their potential energy does not exceed $v$. 
%For this, let us 
%define a trajectory as the set of points resulting of applying all the 
%maps $f^{\nu}$ to a single point in $M_v$. The potential energy of the 
%points in the trajectory, $V^{\nu}({\bf x})=V({\bf 
%x}(\nu))$ is a linear function of $\nu$. Thus, it must be bounded by 
%the potential energy of the endpoints. 
%The initial point, $\nu=0$ has $V({\bf x}) < v$ by definition. The final point 
%is on the sphere $\mathbb{S}^{D-1}$, where
%\begin{equation}
%\frac{V({\bf x} \in \mathbb{S}^{D-1})}{N}=\sum_{k=1}^D \frac{-\lambda_k}{2}\,  
%\frac{x_k^2}{N} \,<\, \sum_{k=1}^D v \,\frac{x_k^2}{N} = v
%\end{equation}
%\noindent using the definition of $v$. We have used the fact that the 
%trajectory is continuous, which depends on the continuity of the map, 
%that can be readily checked. Indeed, Eq.~(\ref{eq.retract}) implies 
%that 
%the map can only be discontinuous in points ${\bf x}_d$ such that 
%$x_i=0$ for $i \leq D$, but these points satisfy
%\begin{equation}
%\frac{V({\bf x}_d)}{N}=\sum_{k=D+1}^N \frac{-\lambda_k}{2}\,  
%\frac{x_k^2}{N} \, \geq \, \sum_{k=D+1}^N v \,\frac{x_k^2}{N} = v
%\end{equation}
%\noindent and thus they are outside $M_v$.

Homotopy equivalence implies that the Betti numbers of the $M_v$ with 
$v\in (v_{j},v_{j+1})$ are the same of $\mathbb{S}^{D-1}$:
$b_i(M_v) =  1$ for $i=0$ and $i=D-1$ and $b_i(M_v) = 0$ otherwise.
Thus at each $v_j$ a {\it topological transition} occurs 
that changes the topology of the phase space from one homotopy 
equivalent to $\mathbb{S}^{D-|C_j|-1}$ to one homotopy equivalent to 
$\mathbb{S}^{D-1}$. In 
terms of the Betti numbers, each transition changes 
two of them from $0$ to $1$ and from $1$ to $0$. Thus, at 
variance with other 
models, the {\it magnitude} of the Betti numbers is not a useful 
quantity in order to characterize the TT. It is better to look at changes 
in $D-1$, the highest {\it index} 
of the Betti number that changes at each transition.
%, i.e. the dimension of the deformation retract of the manifolds. 
Furthermore, as we have shown 
that the manifolds $M_v$ are homotopy equivalent to (hyper)spheres, the 
information about their dimension $D-1$ completely characterizes their topology.
Thus $D$ is the relevant quantity to be studied.
As shown above, the increase 
of $D$ at each TT is given by the degeneracy of the 
corresponding eigenvalue.
%!!!!In Bott's extended Morse theory, we define the {\it order} of the critical
%submanifolds as the number of negative eigenvalues of the Hessian of 
%$V-\mu(\sum_{i=1}^{N} x_i^2-N)$ when restricted to the submanifold normal to the $j$th 
%critical submanifold.!!!! 
If the degeneracy $|C_j|$ is $o(N)$, given that 
$D=\sum_{i=0}^{j} |C_i|$, in 
the $N \rightarrow \infty$ limit $D$ is equivalent to the {\it order} 
of the critical manifolds, which is defined as the number of negative eigenvalues of the Hessian ($\sum_{i=0}^{j-1} |C_i|$)
when restricted to the submanifold normal to the $j$th critical submanifold.
%{\bf Furthermore, the Hessian of the potential has 
%$\sum_{i=0}^{j-1} |C_i|$ negative 
%eigenvalues when restricted to the submanifold 
%normal to the $j$th critical submanifold.
%In particular, if the degeneracy $|C_j|$ is $o(N)$, given that 
%$D=\sum_{i=0}^{j} |C_i|$, in 
%the $N \rightarrow \infty$ limit $D$ is equivalent to the {\it order} 
%of the critical manifolds {\bf ***}.
%, defined as the number of negative 
%eigenvalues of the Hessian of the normal submanifold. 
This generalizes 
to degenerate manifolds the definition of order of a saddle point.

For the spherical model it can be shown that the 
spectrum of eigenvalues is continuous in the infinite $N$ limit. Thus, the set of 
$\hat{N}+1$ critical energies will be dense in 
$[-d,d]$, the interval of allowed potential energies. Consequently
the model has a continuum of TTs. In this limit, and considering that $D$ is 
$O(N)$, 
%Because of this, for infinite size systems 
it is convenient to introduce a 
%use the 
continuous and 
normalized version of $D$, $d(v)=D/N$, and also a {\em degeneracy density},
$c(v)$. They are related by $c(v)=\frac{\partial d(v)}{\partial 
v}$. In the following we search for singularities in these functions or their 
derivatives which could point to particularly strong TTs.

The spectrum of the adjacency matrix is given by~\cite{BK}:

\begin{equation}
\lambda_{\bf p} = 2\sum_{i=1}^d \cos (2 \pi p_i/N^{1/d}) ,\,\,\,\, 
p_i=0, 
\cdots, N^{1/d}-1
\label{eq:espectro}
\end{equation}

\begin{figure}[ht]
\centerline{\epsfig{file=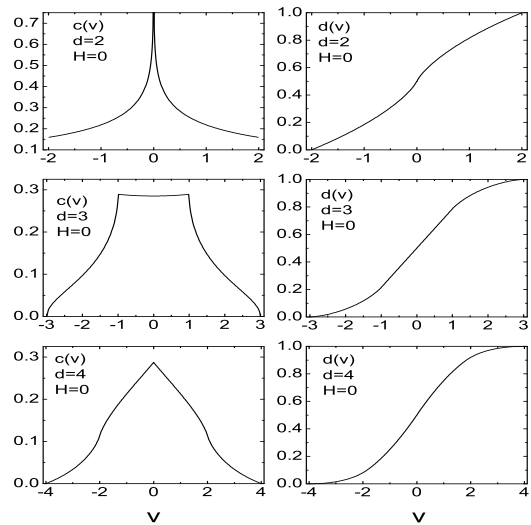,width=7cm,height=7cm}}
\caption{Degeneracy density $c(v)$ and relative 
index $d(v)$ of the critical manifolds, for a vanishing external 
field.} 
\label{fig.d234}
\end{figure}

In the $N\to \infty$ limit, the degeneracy density is:

\begin{eqnarray}
c(v)&=& (2\pi)^d \int_0^{2 \pi} (\Pi_{i=1}^d d \omega_i) \delta (v+ 
\lambda({\bf 
\omega})/2) \nonumber \\
 &=& \int_0^{\infty} \frac{dx}{\pi} \, \cos(x \, v) (J_0 (x))^d
\label{eq:rhoc}
\end{eqnarray}

\noindent where $\lambda({\bf 
\omega})=2\sum_{i=1}^d \cos(\omega_i)$.
It can be shown~\cite{long} 
that the integral converges uniformly 
for all values of $d$ and therefore $c(v)$ is a 
continuous function (see Fig.~\ref{fig.d234}). The 
derivatives with respect to $v$ can be obtained by performing the 
derivative inside the integral. But, as this is only valid if the 
resulting integral converges, this procedure allows us to obtain only the first 
$\lfloor (d-1)/2 \rfloor$ derivatives. All these derivatives are 
continuous~\cite{long} except for the last, which is discontinuous {\it only} at 
the following points: at odd values of $v$ if $d$ is odd, at even values of 
$v/2$ if $d/2$ is odd and at odd values of $v/2$ if $d/2$ is even. But 
these values are clearly {\it different} from the ones at which a 
PT takes place, for all values of $d$ (see Table 1). The manifolds $M_v$ 
display TTs at the points $\langle v_c\rangle$ where PTs occur, 
since there is a continuum of TTs. These TTs, however, are {\it not} particularly
abrupt.
This non coincidence between the levels where a {\it special} TT ($v_{T}$) 
and a PT ($\langle v_c\rangle$)
take place has also been observed in the $\phi^4$ mean field model~\cite{fi4}.

The only possibility left to look for a relationship between 
TT's (in the sense of a discontinuity of some function of the 
topology) and PT's would be in the higher derivatives of $c(v)$, 
which cannot be studied by interchanging the integral and derivative 
operations. This possibility seems 
to us rather unreasonable, because it would imply not only that the 
derivative where discontinuities are to be looked for depends on the 
dimension of the lattice, but also that those discontinuities present 
in lower order derivatives should be disregarded.

We have thus shown that discontinuities in the derivatives of $c(v)$ are 
not sufficient to induce the PT present in the model.
%Nevertheless, it could be expected that if, for 
%some model, discontinuities were present {\it in $c(v)$ 
%itself}, this could be enough to induce a PT. 
%{\bf In the following, adding an external field $H$ to the system, we give a
%counterexample to this possibility.
%Although this model presents no PT, we find discontinuities in its 
%function $c(v)$.}
%that this is not the case.
Furthermore, we show in the following that, even though in the case of a nonzero external field
there appear discontinuities in the function $c(v)$ {\it itself}, no connection between such 
TTs and PTs can exist, simply because the model does not display any PTs at all.
%{\bf Furthermore, we will show in the following that adding an external field
%H the model displays discontinuities in the function $c(v)\,itself$. Nevertheless
%in this case the model does not have a phase transition and consequently no
%connection between this singular TT and a PT can exist.} 

With $H\neq 0$ it is not so easy to find 
the homotopy type of the submanifolds $M_v$, because of the 
breaking of the symmetry introduced by the 
field term in the Hamiltonian. 
Nevertheless, using Morse theory it is at least  possible to establish 
the homotopy type of the submanifolds up to above the second smallest 
critical energy, where an abrupt topological change is shown to take place.

According to Morse theory, if there is one nondegenerate critical point at 
$c \in (a,b)$, the 
manifold $M_{c+\epsilon}$ is homeomorphic to $M_{c-\epsilon} \cup\, 
e_{k}$, where $e_{k}$ is a $k$-cell. In other words, at the critical point, a 
$k$-cell (i.e. a $k$-dimensional open disk) is attached to the 
manifold, where
$k$ is the {\it index} of the critical point, defined as the number of 
negative eigenvalues of the Hessian at that point.

From the critical point equations (\ref{eq.critp}) we obtain that 
the smallest critical energy is $v_+=-(\lambda_1/2+H)$, and the next 
is $v_-=-(\lambda_1/2-H)$. The 
Hessian of the 
potential on the sphere at the critical points 
${\bf x}_{\pm}=(\pm \sqrt{N},0, \cdots ,0)$ is a diagonal matrix 
with $V^{\pm}_{ii}=\lambda_1 \pm H -\lambda_i$ ($i > 1$). Therefore, at 
these two 
points the Hessian is not singular, which implies that ${\bf x}_{\pm}$ are nondegenerate critical 
points. But $\lambda_1$ is the largest eigenvalue, therefore for $v_+$ 
all 
the eigenvalues $V_{ii}$ are positive. This was to be expected 
because this is 
the absolute minimum of the potential. Topologically this means that 
for $v< v_-$ $M_v$ is homotopy equivalent to a disk.
For $v=v_-$ the index of the critical point depends on the field: if 
$\lambda_2 <  \lambda_1-H$, $v_-$ is a minimum. 
Thus, denoting the next critical value by $v_2$,
%if $v_2$ is the next critical point, 
$M_v$ for $v \in (v_-,v_2)$ is homotopy equivalent to the union of 
two disjoint disks on the sphere.
However, for large values of $N$ the topological scenario is diferent. 
Since in this limit the spectrum of the 
adjacency matrix becomes dense, a certain number $k$ 
of its eigenvalues will fall into the interval
$(v_+,v_-)$.
This number becomes the {\it order} of the critical point at $v_-$, and gives 
the dimension of the $k$-cell that is attached to the disk. The 
manifold $M_v$ for $v \in (v_-,v_2)$ is therefore homotopy 
equivalent to 
a sphere of $k$ dimensions. For large $N$, $k$ 
becomes proportional to $N$. In the 
interval $(v_+,v_-)$ the manifolds $M_v$ have the homotopy type of a 
point. At $v_-$ an abrupt change in the 
topology takes place, and the 
$M_v$ have now the homotopy type of a sphere with a macroscopic 
number of dimensions (see the jump of $c(v)$ in Fig. 2). 

\begin{figure}[ht]
  \centerline{\epsfig{file=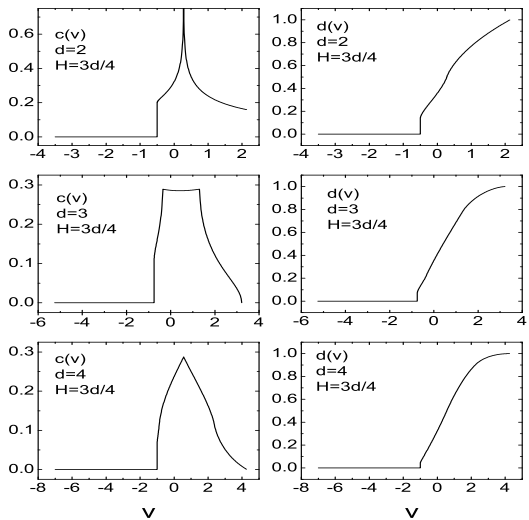,width=7cm,height=7cm}}
  \caption{Degeneracy density $c(v)$ and relative 
index $d(v)$ of the critical manifolds, for a finite field. A delta 
function is assumed at the discontinuity point in the graphs of the 
left column.} \label{fig.d234H}
\end{figure}

For higher values of $V$, the critical values are given by 
$v_j=-\lambda_j/2+H^2/2(\lambda_1-\lambda_j)$, but only for $j$ such 
that $\lambda_1 - \lambda_j  > H$. 
%{\bf This is consistent with the absence 
%of critical values between $v_+$ and $v_-$.}
%{\bf The critical values $v_j$ correspond 
%to critical {\it submanifolds}, given by $\{\mathbf x \in \Gamma : x_1=\sqrt{N}H/(\lambda_j-\lambda_1), 
%\sum_{k\in C_j}x_k^2=1-H^2/(\lambda_j-\lambda_1)^2\}.$}
Notice that, at variance with the case of 
vanishing $H$, there is a threshold energy below which 
the critical values have been suppressed. 
The critical submanifolds occuring at each $v_j$ are again hyperspheres 
whose dimension is given by the degeneracy of the corresponding 
eigenvalue $\lambda_j$. For all critical values we have calculated the order of the 
critical manifolds, $d(v)$, as well as the relative degeneracy, 
$c(v)$, for a few values of $d$ (see Fig. 2).
The main difference with the results for $H=0$ is 
that now the connection between the order and the topology of 
the different manifolds is less obvious, and we have not been able 
to identify the homotopy types for all values of $v$. Nevertheless we have 
found exactly the topological change 
that takes place at $v_-$, and have shown that it is {\it 
macroscopic}. It 
may come as a surprise that this very abrupt change does not have a 
PT associated to it. 

%The use of tools from algebraic topology allowed us to characterize 
%exactly the homology of the successive manifolds of the configuration 
%space of the short range spherical model. We have shown that even 
%though there is a 
%continuum of topological 
%transitions, a function of the topology can be defined whose 
%derivatives display some discontinuities which could be associated to PTs. 
%For vanishing field, however, 
%they are not sufficient to induce a PT. We also explored the 
%possibility that
%discontinuities in the function itself could be able to induce a PT. 
%For the case of nonvanishing 
%field we have shown that, even though such discontinuities are 
%present, they do not have an associated PT. These results seem to be
%against the ubiquity of the topological hypothesis, at least in its
%present form.

We have shown that the manifolds of the configuration 
space of the short range spherical model
display a continuum of topological transitions. Hence the necessity
condition implied by the theorem in \cite{FP} is trivially met.
Also, strong discontinuities
have been found either in a function of the topology or in its derivatives.
Although these discontinuities represent abrupt changes in the topology 
we have shown that they are not associated to PTs. 
Conversely, at the points where PTs take place no
abrupt changes are observed in the topology. These are the first results 
on a short range confining potential to challenge the
sufficiency of a topological mechanism in the origin of a phase transition.
%as proposed by the topological hypothesis in its present form.

One intriguing question that arises is whether the topologically abrupt 
changes that we have found can have some influence on the {\it 
dynamics} of the model.

This work was partly supported by CNPq (Brazil). S. R-G. acknowledges 
support from the Centro Latinoamericano de Física.

\end{document}